\documentclass[pra,
               twocolumn,
               a4paper,amsmath,twocolumn,10pt]{revtex4-2} 
\usepackage[utf8]{inputenc}
\usepackage[]{hyperref} 
\usepackage{url}

\usepackage{graphicx}
\usepackage{latexsym}
\usepackage{amssymb}
\usepackage{amsmath}
\usepackage{graphics}
\usepackage{layout}
\usepackage{verbatim}
\usepackage{amsfonts,epsfig,dsfont}
\usepackage{amssymb,amsmath}
\usepackage{mathtools}
\usepackage{color}
\usepackage{orcidlink}

\hypersetup{pdfauthor={Igor Bragar},%
            pdftitle={Retardation of Entanglement Decay of Two Spin Qubits by Quantum Measurements},%
            pdfsubject={Entanglement of spin qubits},%
            pdfkeywords={spin, qubit, entanglement, concurrence, negativity, quantum dot, quantum measurement, projective measurement},%
}

\newcommand{\h}{\hat}
\newcommand{\beq}{\begin{equation}}
\newcommand{\eeq}{\end  {equation}}
\newcommand{\tr}[1]{\mathrm{Tr} \left( {#1} \right)}

\makeatletter
\newcommand{\bal}{\@ifstar{\@bals}{\@bal}}
\def\@bals#1\eal{\begin{align*}#1\end{align*}}
\def\@bal#1\eal{\begin{align}#1\end{align}}
\makeatletter

\newcommand{\eqcolon}{\mathrel{\resizebox{\widthof{$\mathord{=}$}}{\height}{ $\!\!=\!\!\resizebox{1.2\width}{0.8\height}{\raisebox{0.23ex}{$\mathop{:}$}}\!\!$ }}}

\DeclareUnicodeCharacter{2009}{\,} 

\begin{document}

\title{Retardation of Entanglement Decay of Two Spin Qubits 
       by~Quantum Measurements 
      }
\author{Igor Bragar\,\orcidlink{0000-0003-1321-2371}}
\email{Igor.Bragar@gmail.com}
\affiliation{Institute of Physics,
             Polish Academy of Sciences,
             al.~Lotnik{\'o}w 32/46,
             02-668 Warszawa, Poland
            }
\date{\today}
\begin{abstract}
We study a system of two electron spins 
each interacting with its small nuclear spin environment (NSE), 
which is a prototype system of two electron spin quantum dot (QD) qubits.
We propose a way to counteract the decay of entanglement in two-electron spin subsystem (TESSS) 
by performing some manipulations on TESSS (the subsystem to which experimentalists have an access), 
e.g. repeatable quantum projective measurements of TESSS. 
Unlike in the quantum Zeno effect, 
the goal of the proposed manipulations is not to freeze TESSS in its initial state 
and to preclude any time evolution of the state by infinitely frequent quantum measurements. 
Instead of that, 
performing a few cycles of free evolution of the system for some time $\tau$ 
followed by a quantum measurement of TESSS 
with subsequent postselection of TESSS state (the same as the initial one) 
produces quantum correlations in NSEs 
and also restores the quantum correlations in TESSS. 
By numerical calculation of the system evolution (the full density matrix $\hat \rho(t)$), 
we show 
that,
in contrast to the fast decay of TESSS entanglement 
on timescale of the order of $T_2^*$,
application of the proposed manipulation sequence
gradually builds up coherences in the entire system 
and the rest decay of quantum correlations of TESSS 
may be significantly slowed down for specific cycle durations $\tau$ 
and numbers of performed cycles.
\end{abstract}

\maketitle

\section{Introduction} 
\label{sec:Introduction} 
Spin of an electron localized on a quantum dot (QD) in a semiconductor nanostructure
is a promising physical realization of qubit 
as it can be reliably initialized, manipulated, and read out
\cite{Kim_NatPhys_2011,DeGreve_RepProgPhys_2013}.
To be a useful element of a quantum computer,
such a qubit must fulfill, among others,
the criterion of long decoherence times
\cite{Loss_PhysRevA_1998,DiVincenzo_FdP_2000}.
Providing no manipulations aimed at mitigation of the influence 
exerted by the environment on the spin qubits have been applied, 
coherences as well as quantum correlations of a pair of electron spin QD qubits decay
on a nanosecond timescale
\cite{Cywinski_APPA_2011}.
The main factor of such a fast decay is 
the Fermi contact hyperfine interaction of electron spin 
with nuclear spins of atoms 
from which the nanostructure is built
\cite{Mazurek_PhysRevA_2014,Bragar_PhysRevB_2015}.

There have already been proposed and implemented in the experiment a few strategies
to enhance the decoherence times of electron spin
such as:
dynamical decoupling of spin qubits from their environments
\cite{Bluhm_NatPhys_2011};
preparing an artificial state of environment
(so called narrowed state of nuclear spin bath)
\cite{Bluhm_PhysRevLett_2010};
or simply making use of 
materials which are made of atoms with spinless nuclei, e.g. isotopically-purified $^{28}$Si
\cite{Tyryshkin_NatMater_2012}.
All these strategies can be summarized as: avoiding as much as possible 
any interaction of the qubits with their environments.

In this paper we propose the opposite strategy 
to counteract the decoherence 
and induced by it decay of quantum correlations
of two electron spin qubits. 
We explore the process of transfer of coherences and quantum correlations 
from a pair of entangled qubits to the environment 
combined with quantum measurements of the qubits' subsytem.
Provided that the environment is non-Markovian, 
e.g. preserves some memory of past interactions,
it turns out that it, being in a quantum state 
obtained after a period of free evolution of the system,
can dephase qubits with a lower rate.
Using a simple model of a system of two electron spin QD qubits, 
presented in Sec.~\ref{sec:Model},
we investigate the effect of application of the manipulation procedure 
described in Sec.~\ref{sec:Manipulation}
on dynamics of entanglement decay.
Results are discussed in Sec.~\ref{sec:Results}, 
where it is shown that both parts of the procedure, 
namely free evolution of the system
and quantum measurement of the qubits' subsystem 
with subsequent postselection of the two-qubit quantum state,
are equally important, 
and that only for specific combinations of durations of free evolution periods $\tau$ 
and number of cycles $n$, 
significant retardation of entanglement decay can be achieved.

We would like to stress 
that the proposed manipulation procedure is not a realization 
of the quantum Zeno effect 
\cite{Misra_JMatPhys_1977}. 
Here, 
the goal is not to freeze qubits in their initial state
and to preclude any time evolution of the state by infinitely frequent quantum measurements. 
Instead of that, 
we let qubits interact with their environments 
and transfer to them some coherences and quantum correlations
during joint system evolution.

\section{The Model of Electron Spin Quantum Dot Qubits} 
\label{sec:Model} 
First, we describe the model of electron spin QD qubits, 
which we use to illustrate the proposed manipulation sequence.
We consider a system of two semiconductor QDs
(e.g. self-assembled InGaAs QDs or gated QDs created in GaAs-AlGaAs nanostructure), 
each of which has a localized electron on it.
As such systems usually are operated at low temperatures,
we suppose that electrons are in their ground states.
In such a case
one can exclude from further consideration
the spatial part of the electron's wave functions 
and focus only on the spin parts of the wave functions.

For the sake of clarity,
we also assume that 
during periods of free evolution of the system
there is no any inter-QD interaction,
which could create some entanglement between the two QDs 
and, especially, between electron spins 
(e.g. electrons are strongly localized on QDs
because of a high enough inter-QD potential barrier 
or relatively long distance between the QDs and
the electron's wave functions hardly overlap, 
so no interaction between the two electron spins occurs).
It is worth to mention 
that we will be analyzing the behaviour of entanglement 
on the time scale from 0 to a few $T_2^*$. 
For such short times
none realistic part of the interaction Hamiltonian, 
apart from the Fermi contact hyperfine interaction of electron spin 
with nuclear spins from its environment, is essential, 
because it does not manifest at short times 
(the energies of dipolar or quadrupolar interaction of nuclear spins
are orders of magnitude lower than 
energy of hyperfine interaction),
whereas the Fermi contact hyperfine interaction leads to the fast complete decay
of initially present in the electron spins subsystem entanglement 
in any finite magnetic field \cite{Bragar_PhysRevB_2015}.

Thus, the Hamiltonian of the system of two QDs has the form:
\beq
\hat H = \hat H^{(1)}   \otimes \mathds{1}
       + \mathds{1} \otimes \hat H^{(2)}.
 \label{eq:SystemHamiltonian}
\eeq
The Hamiltonian $\hat H^{(i)}$ of a single QD contains the following terms
\beq
\hat H^{(i)} = \hat H_{\mathrm{el.}}^{(i)}
             + \hat H_{\mathrm{nucl.~env.}}^{(i)}
             + \hat H_{\mathrm{int.}}^{(i)}.
 \label{eq:QDHamiltonian}
\eeq
The first and the second terms of $\hat H^{(i)}$ are 
the Zeeman energies of electron and nuclear spins, respectively:
\beq
\hat H_{\mathrm{el.}} = \Omega \hat S_{z} \otimes \mathds{1}^{\otimes N},
 \label{eq:ElectronHamiltonian}
\eeq
where  $\Omega = g_{\mathrm{eff.}} \mu_{\mathrm{B}} B_z$ 
is the Zeeman splitting of electron spin,
$g_{\mathrm{eff.}}$ is effective 
g-factor of electron spin,
$\mu_{\mathrm{B}}$ is the Bohr magneton,
$B_z$ is $z$-component of magnetic field,
$\hat S_z$ is the operator of the $z$-component of electron spin,
$N$ is the number of nuclear spins interacting with electron spin.
For the sake of simplicity, 
we have also adopted the assumption 
that all nuclear spins are of the same type $J$,
so the identity operator $\mathds{1}$ used above is of dimension $2J + 1$.
\beq
\hat H_{\mathrm{nucl.~env.}} = \sum_{n = 1}^{N} \omega^{(n)} 
                                       \mathds{1}^{\otimes (n-1)} 
                               \otimes \hat J_{z}^{(n)} 
                               \otimes \mathds{1}^{\otimes (N-n)},
 \label{eq:NuclearEnvironmentHamiltonian}
\eeq
where $\omega^{(n)} = g^{(n)} \mu_{\mathrm{N}} B_z$
is the Zeeman splitting of $n$th nuclear spin,
$g^{(n)}$ in the nuclear g-factor of $n$th nuclear spin,
$\mu_{\mathrm{N}}$ is the nuclear magneton.

The last term of the Hamiltonian $\hat H^{(i)}$ is the hyperfine interaction 
between electron spin and nuclear spins:
\bal
\hat H_{\mathrm{int.}} &= \sum_{n = 1}^{N} A_{n} \hat {\mathbf{S}}
                          \otimes \mathds{1}^{\otimes (n-1)} 
                          \otimes \hat {\mathbf{J}}^{(n)}
                          \otimes \mathds{1}^{\otimes (N-n)},
\eal
where $\hat {\mathbf{S}} = \left(\hat S_x, \hat S_y, \hat S_z\right)$ is the electron spin operator,
$\hat {\mathbf{J}}^{(n)} = \left(\hat J_x^{(n)}, \hat J_y^{(n)}, \hat J_z^{(n)} \right)$
is the $n$th nuclear spin operator
and 
$A_n$ is the hyperfine coupling between electron spin and $n$th nuclear spin.

\section{Manipulation Procedure with Quantum Measurements 
         and Postselection of Two-Qubit State}            
\label{sec:Manipulation} 
Motivated by experimentalists' capabilities 
to initialize localized in QDs electrons in singlet state 
and to perform projective measurements onto singlet state
\cite{Kim_NatPhys_2011,DeGreve_RepProgPhys_2013}, 
we consider a quantum measurement of two-electron spin subsystem 
(TESSS),
specifically, the measurement answering whether TESSS is in 
a certain two-qubit state or not.
In general, such a~quantum measurement can be described by measurement operators
$\h M_1$ (``yes'' result) and $\h M_2$ (``no'' result):
\bal
\hat M_1 &= \sqrt{k}   \, \hat \Pi_{\mathrm{2q}} \otimes \mathds{1}_{\mathrm{2env}} 
    + \sqrt{1-k} \, \left(\mathds{1} - \hat \Pi_{\mathrm{2q}} \otimes \mathds{1}_{\mathrm{2env}} \right), \\
\hat M_2 &= \sqrt{1-k} \, \hat \Pi_{\mathrm{2q}} \otimes \mathds{1}_{\mathrm{2env}} 
    + \sqrt{k}   \, \left(\mathds{1} - \hat \Pi_{\mathrm{2q}} \otimes \mathds{1}_{\mathrm{2env}} \right).
\eal
where 
$\hat\Pi_{\mathrm{2q}}$ is a projector in TESSS subspace onto a chosen two-qubit state,
parameter  $k  \! \in \! [\frac{1}{2}, 1]$ is a strength of measurement,
$\mathds{1}$ is the identity operator of dimension of the system state space,
and
$\mathds{1}_{\mathrm{2env}}$ is the identity operator 
of dimension of the two NSEs subsytem state space.
The extreme values of the quantum measurement strength have clear physical meanings:
$k = 1$ corresponds to the case of measurements of the highest strength, 
i.e. the projective measurement,
\bal
\hat M_1 &= \hat \Pi_{\mathrm{2q}} \otimes \mathds{1}_{\mathrm{2env}}, \\
\hat M_2 &= \mathds{1} - \hat \Pi_{\mathrm{2q}} \otimes \mathds{1}_{\mathrm{2env}},
\eal
and $k = \frac{1}{2}$ corresponds to the case of completely ineffective measurement,
\beq
\hat M_1 = \hat M_2 = \frac{1}{\sqrt{2}} \mathds{1}.
\eeq
The intermediate values of strength $k$ correspond to such quantum measurements
that give the outcomes
which are the probabilistic mixture of the outcomes of projective operators
$\hat \Pi_{\mathrm{2q}} \otimes \mathds{1}_{\mathrm{2env}}$
and $\mathds{1} - \hat \Pi_{\mathrm{2q}} \otimes \mathds{1}_{\mathrm{2env}}$,
i.e. the fidelity of the outcomes, 
compared with these of projective measurement,
is determined by the measurement strength 
and is equal to $k^2$.
By construction, the measurement operators $\h M_1, ~\h M_2$ fulfill the completeness relation
$\sum_{i = 1}^{2} \hat M_i^{\dagger} \hat M_i \equiv \mathds{1}$ for any $k$ from its range.

\begin{figure}[t!]
  \centering
  \includegraphics[width=\linewidth, viewport=0 0 692 140, clip]
  {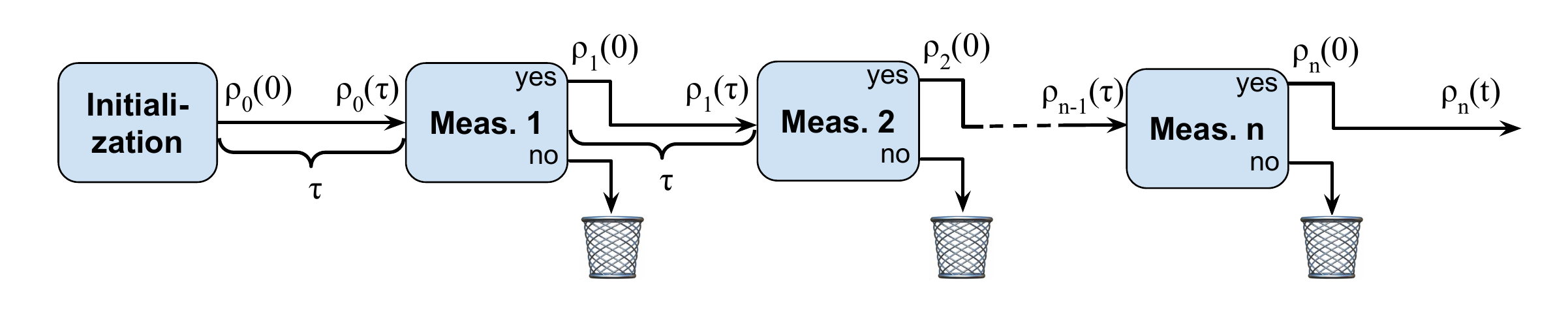}
  \caption{
            Schematic representation 
            of the proposed manipulation sequence 
            with quantum measurements and postselection of the two-qubit state.
          }
  \label{fig:manipulation_sequence}
\end{figure}

The manipulation procedure consists of
initialization of the system in its initial state
and a few manipulation cycles.
The manipulation cycle, in turn, has two parts: 
free evolution of the system for a time $\tau$
and execution of quantum measurement 
with subsequent postselection of TESSS state.
The idea of the manipulation procedure is shown in Fig.~\ref{fig:manipulation_sequence}.
First, the system is initialized in a state 
$\hat \rho_{\mathrm{ini}} = { \hat \rho_{\mathrm{2q}}(0) \otimes \hat \rho_{\mathrm{2env}}(0) }
\eqcolon \hat \rho_0 (0)$.
The initial TESSS state, 
$\hat \rho_{\mathrm{2q}}(0)$, 
is supposed to be a maximally entangled state,
whereas NSEs usually are in a high-temperature state \cite{Cywinski_APPA_2011}, 
which has no coherences at all,
if no manipulations have been performed on them beforehand.
Such a choice of NSE initial state is physically motivated: 
even low temperatures of the order of a few hundred of mK, 
at which experiments with QDs are routinely performed, 
are already sufficiently high enough for nuclear spins
due to the smallness of their Zeeman energies or dipolar interaction 
compared to $k_{\mathrm{B}}T$.
Thus, the initial state of NSEs has the form 
$\hat \rho_{\mathrm{2env}}(0) = \hat \rho_{\mathrm{env_1}}(0) \otimes \hat \rho_{\mathrm{env_2}}(0)$,
where $\hat \rho_{\mathrm{env_i}}(0) = \frac{1}{Z_i} \mathds{1}$, 
$Z_i = (2J+1)^{N_i}$, $i = 1,2$.

After initialization, 
we let the system freely evolve for some time $\tau$
obtaining the state 
$\hat \rho_0(\tau) = \hat U(\tau) \h \rho_0 (0) \hat U^{\dagger}(\tau)$,
where $\hat U(\tau) \coloneqq \exp\left(-\frac{i}{\hbar} \hat H \tau \right)$.
Next, a quantum measurement of TESSS is performed 
producing, according to the measurement postulate of quantum mechanics 
\cite{Nielsen_book_QCQI}, 
in an indeterministic way,
one of two possible states
\beq
\hat \rho_1^{\textrm{yes}}(0) \coloneqq 
\hat M_1 \hat \rho_{0} (\tau) \hat M_1^{\dagger} / \tr {\hat M_1 \hat \rho_{0} (\tau) \hat M_1^{\dagger}}
\eeq
or
\beq
\hat \rho_1^{\textrm{no}}(0) \coloneqq  
\hat M_2 \hat \rho_{0} (\tau) \hat M_2^{\dagger} / \tr {\hat M_2 \hat \rho_{0} (\tau) \hat M_2^{\dagger}}
\eeq
with probabilities calculated according to the Born's rule 
$p^{\textrm{yes}} = \tr {\hat M_1 \hat \rho_{0} (\tau) \hat M_1^{\dagger}}$
and
$p^{\textrm{no}} = \tr {\hat M_2 \hat \rho_{0} (\tau) \hat M_2^{\dagger}}$,
respectively.

The state $\hat \rho_1^{\mathrm{yes}}(0)$, which corresponds to the operator $\hat M_1$,
is postselected for further manipulations.
If the outcome of the measurement happens to be the state $\hat \rho_1^{\mathrm{no}}(0)$,
then it is rejected and  execution of the manipulation procedure is interrupted.
After successful execution of the $n$th manipulation cycle, 
the state 
\beq
\h \rho_{n} (0) \coloneqq {     \h \rho_{n}^{\h M_1} (0)} / {\tr{\h \rho_{n}^{\h M_1} (0)}},
\eeq 
where
$\h \rho_{n}^{\h M_1} (0) \coloneqq \h M_1 \h \rho_{n-1} (\tau) \h M_1^{\dagger}$,
is obtained,
for which we study dynamics of TESSS entanglement.

\section{Results and Discussion} 
\label{sec:Results}
We would like to note 
that it is crucial in the simulations to keep the density matrix of the whole system, 
$\hat \rho_n (t)$,
and not to reduce it to the two-qubit density matrix $\hat \rho_{\mathrm{2q}}(t)$
by tracing out NSEs.
Having at hand the full density matrix,
one can investigate the transfer of coherences and quantum correlations in the system 
to the greatest degree.
As dimension of the system state space grows exponentially with the number of nuclear spins,
our capabilities to simulate application of the proposed manipulation procedure
are limited to small systems,
so 
we present here the results obtained for 
the system of two QDs with homonuclear ($J = \frac{1}{2}$) NSEs
of the same size $N_1 = N_2 = 5$.

In simulations, as an initial TESSS state we have used singlet state,
$\hat \rho_{\mathrm{2q}}(0) = \left| \psi_- \rangle \langle \psi_- \right |$,
where 
$| \psi_- \rangle = 
\frac{1}{\sqrt 2} 
\left(   \left| \uparrow \downarrow \right. \rangle 
       - \left| \downarrow \uparrow \right. \rangle \right)$.
The projector operator $\h \Pi_{\mathrm{2q}}$ 
have also been chosed to be a projector onto singlet state,
$\h \Pi_{\mathrm{2q}} \! = \! | \psi_- \rangle \langle \psi_- |$.

To quantify the amount of entanglement of TESSS state $\hat\rho$,
we use concurrence \cite{Wootters_PhysRevLett_1998}, 
which is defined as 
\begin{equation}
 C(\hat\rho) = \mathrm{max}\left\{ 0, \lambda_1-\lambda_2-\lambda_3-\lambda_4\right\} \, ,
\end{equation}
where $\lambda_1 \geqslant \lambda_2 \geqslant \lambda_3 \geqslant \lambda_4$
are the square roots of the eigenvalues of matrix $\hat \rho \hat{\tilde{\rho}}$, 
where $\hat{\tilde{\rho}} = (\hat\sigma_y \otimes \hat\sigma_y ) \hat \rho^* (\hat\sigma_y \otimes \hat\sigma_y )$.
Here ${\hat\rho}^*$ denotes the operation of complex conjugation
of each element of $\hat\rho$.
The concurrence ranges from $C=0$ for separable states
to $C=1$ for maximally entangled states.
We also use negativity \cite{Vidal_PhysRevA_2002} 
to estimate the level of entanglement between two parts of the system.
We show below that of particular interest is the negativity between TESSS and NSEs.

While considering the quantum correlation dynamics, 
it is convenient for further analysis 
to express time in units of two-qubit $T_2^*$
defined as follows
\beq
\frac{1}{\Big(T_2^*\Big)^2} \! = \! \frac{1}{\Big(T_2^{*(1)}\Big)^2} 
                       +    \frac{1}{\Big(T_2^{*(2)}\Big)^2}, 
\eeq
where $T_2^{*(i)}$ is the single-qubit dephasing time, 
$T_2^{*(i)} \! = \! \hbar\sqrt{8 N_{i}}/ \sum_{n=1}^{N_{i}} A_n^{(i)}$.
The decay of entanglement of two electron spin qubits 
plotted using such a time unit
is independent of the system size 
and the absolute value of the hyperfine interaction
\cite{Bragar_PhysRevB_2015}.

\begin{figure}[t!]
  \centering
  \includegraphics[width=\linewidth, viewport=0 0 521 579, clip]{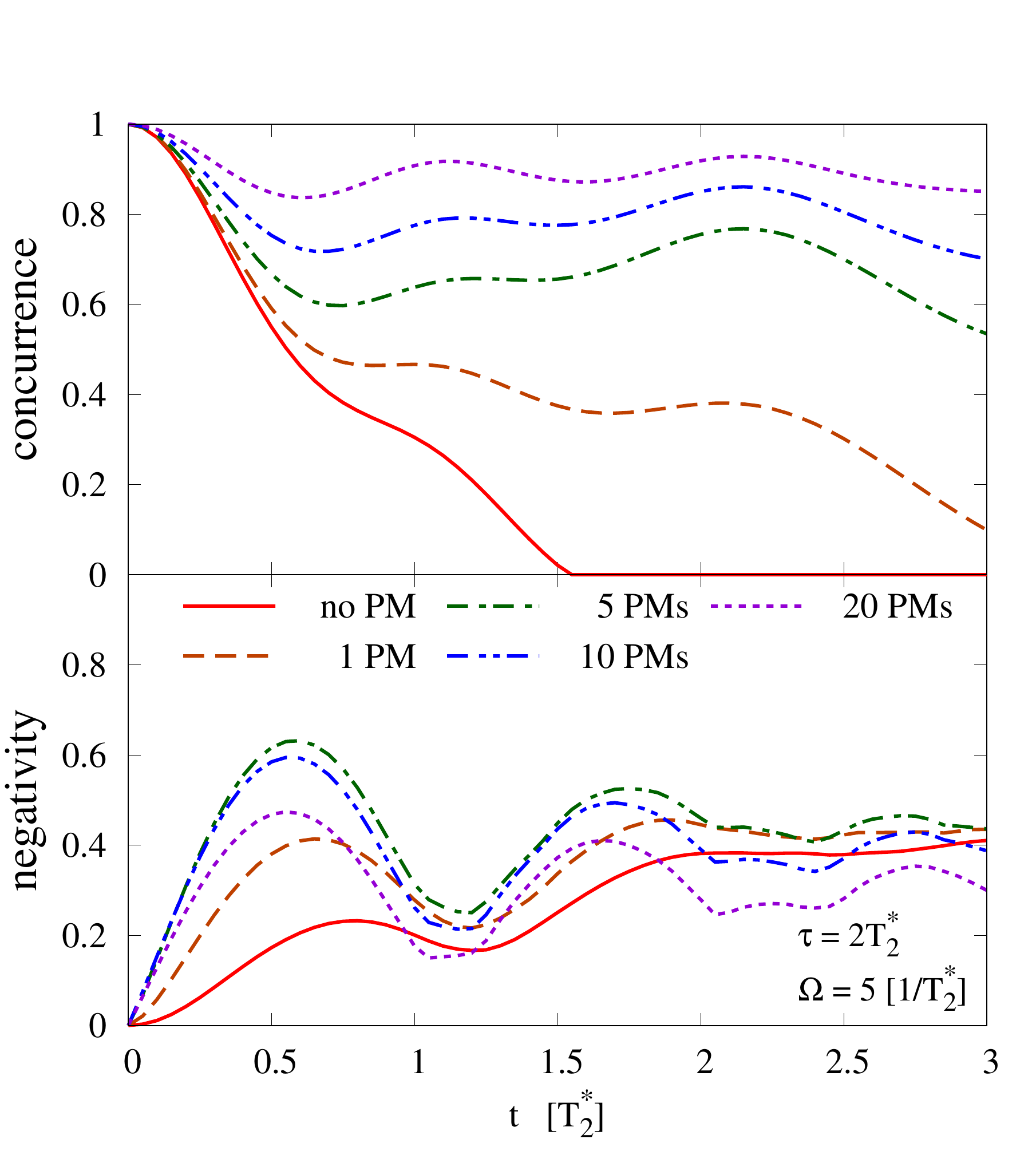}
  \caption{
            Concurrence of two-qubit state $\h \rho_{\mathrm{2q}}(t)$
            and negativity calculated for the system state $\h\rho_n(t)$
            divided into two parts, TESSS and NSEs,
            as functions of time $t$ after the last projective measurement (PM).
            NSEs consist of $N_1=N_2=5$ uniformly coupled  spins~$\frac{1}{2}$. 
            The system is in moderate magnetic field 
            $\Omega = 5 \left[ \frac{\hbar}{T_2^*} \right]$.
          }
  \label{fig:concurrence_negativity_plot}
\end{figure}

The results of simulations,
which are shown in figures 
\ref{fig:concurrence_negativity_plot}--\ref{fig:concurrence_k_plot.pdf},
have been obtained for 
the system being in moderate magnetic field 
$\Omega = 5 \left[ \frac{\hbar}{T_2^*} \right]$.
In Fig.~\ref{fig:concurrence_negativity_plot} 
the time dependencies of concurrence of TESSS state (top panel)
and negativity (bottom panel) are shown 
for a few different numbers $n$ of performed cycles.
As it can be seen from the top panel of Fig.~\ref{fig:concurrence_negativity_plot},
normally entanglement is completely lost after time $t \approx 1.5 T_2^*$, 
but application of just a single manipulation cycle causes a significantly rise
of the entanglement level at all times and noticeably retards its decay.
With increasing number $n$ of performed cycles,
level of entanglement systematically grows, reaching almost its maximal value.
Along with the decay of entanglement in TESSS, 
we see appearance of entanglement between initially uncorrelated parts of the system, 
namely between TESSS and their NSEs 
(see bottom panel of Fig.~\ref{fig:concurrence_negativity_plot}).

\begin{figure}[t]
  \centering
  \includegraphics[width=\linewidth, viewport=0 0 432 360, clip]{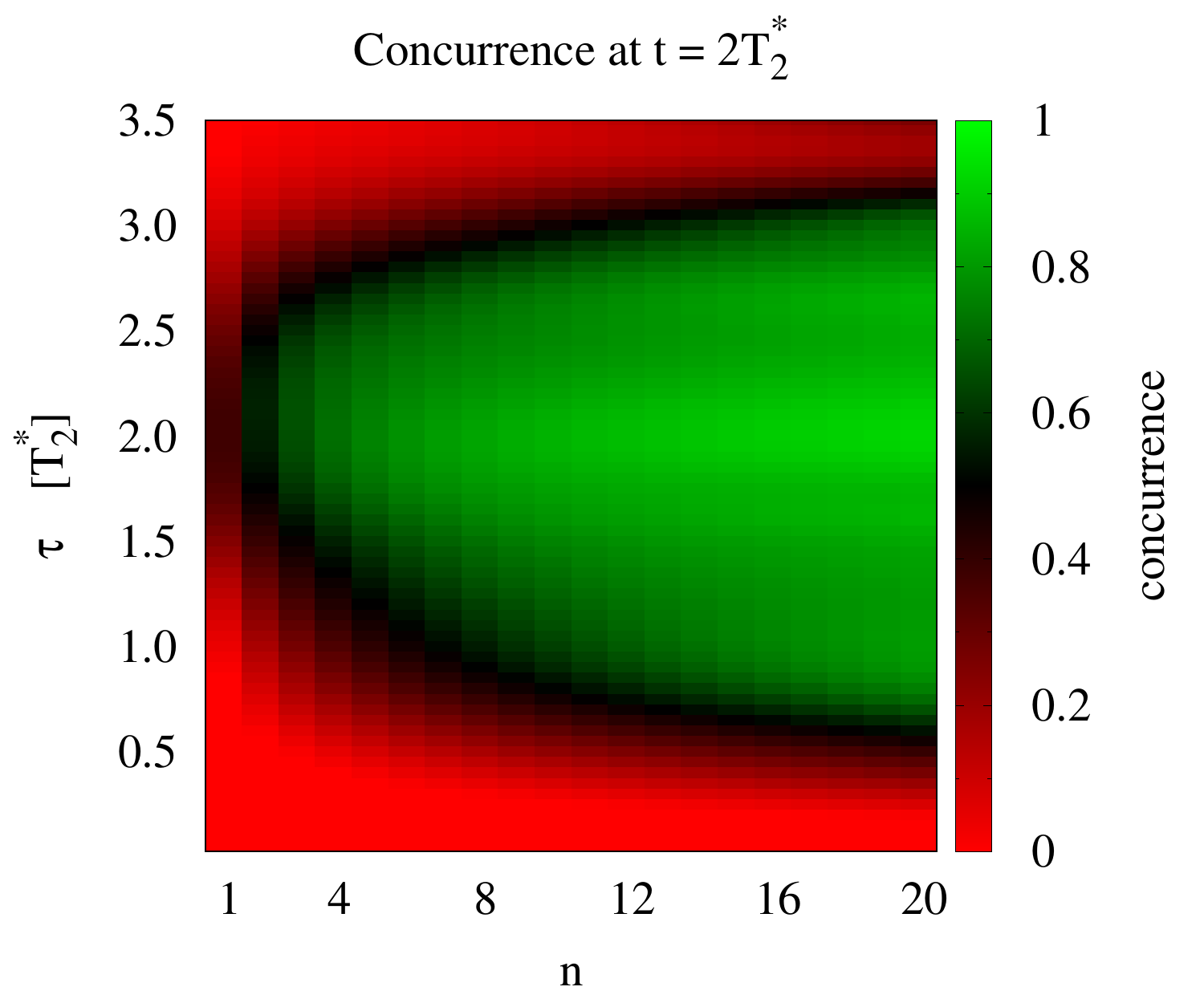}
  \caption{
            Concurrence calculated for $\hat \rho_{\mathrm{2q}}(t = 2T_2^*)$ 
            as a function of number $n$ of performed projective measurements
            and time~$\tau$ between them.
            NSEs consist of $N_1=N_2=5$ uniformly coupled spins~$\frac{1}{2}$. 
            The system is in moderate magnetic field 
            $\Omega = 5 \left[ \frac{\hbar}{T_2^*} \right]$.
          }
  \label{fig:concurrence_map}
\end{figure}

In order to estimate the effect of retardation of entanglement decay 
produced by application of the manipulation procedure,
we monitor the level of concurrence calculated for $t = 2 T_2^*$,
which is shown
in Fig.~\ref{fig:concurrence_map},
as a function of number $n$ of performed projective measurements 
and time $\tau$ between them.
Using this map, one can find the optimal values of the parameters 
$n$ and $\tau$, 
which maximize the retardation of entanglement decay.
On one hand, 
increasing the number of manipulation cycles almost always 
enhances
the effect,
on the other hand, 
it turns out that there exists the most optimal duration $\tau$ 
of the free evolution periods ($\tau_{\mathrm{opt.}} \approx 2 T_2^*$ for the simulated system).

\begin{figure}[t!]
  \centering
  \includegraphics[width=\linewidth, viewport=0 0 504 360, clip]{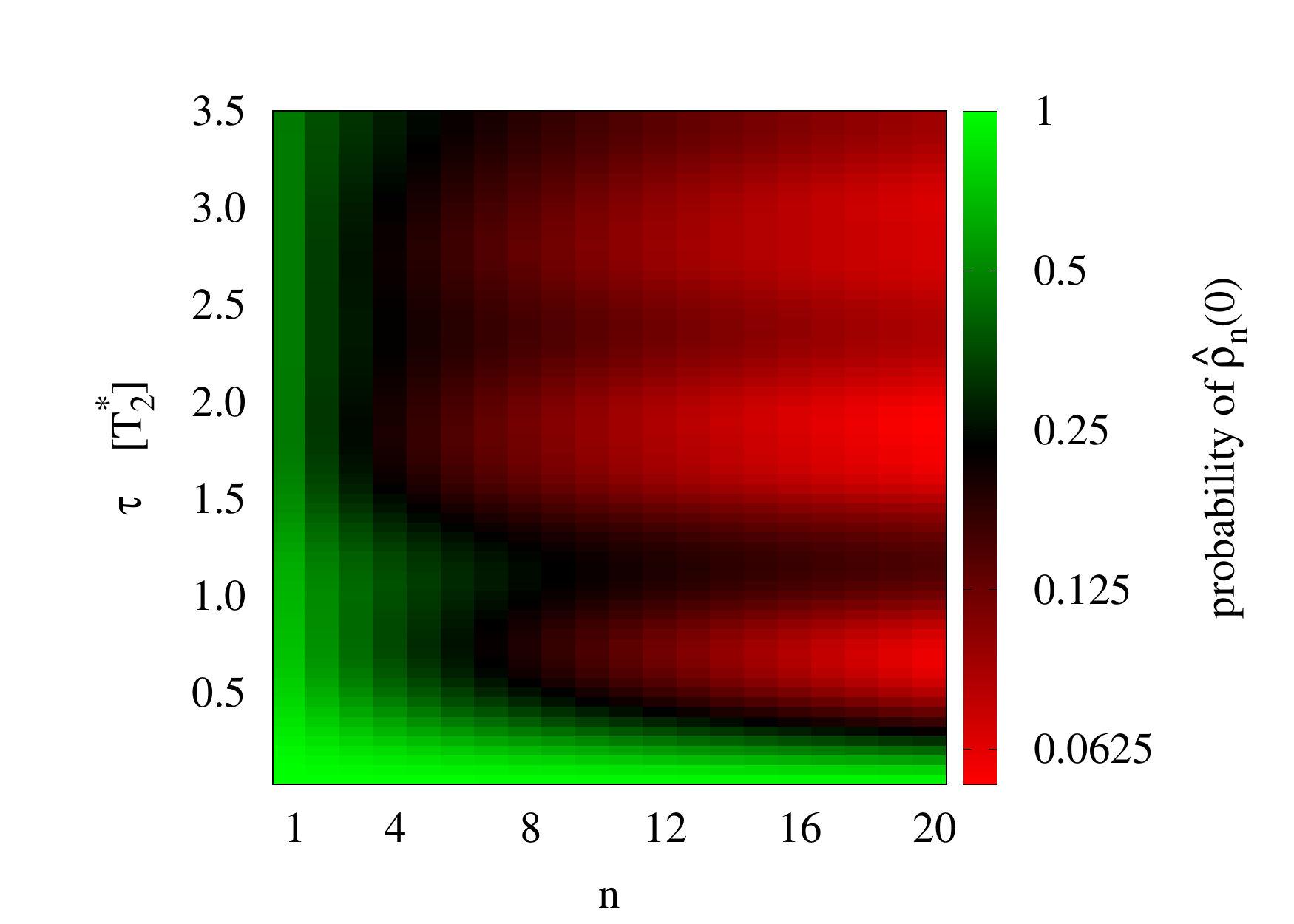}
  \includegraphics[width=\linewidth, viewport=0 0 719 277, clip]{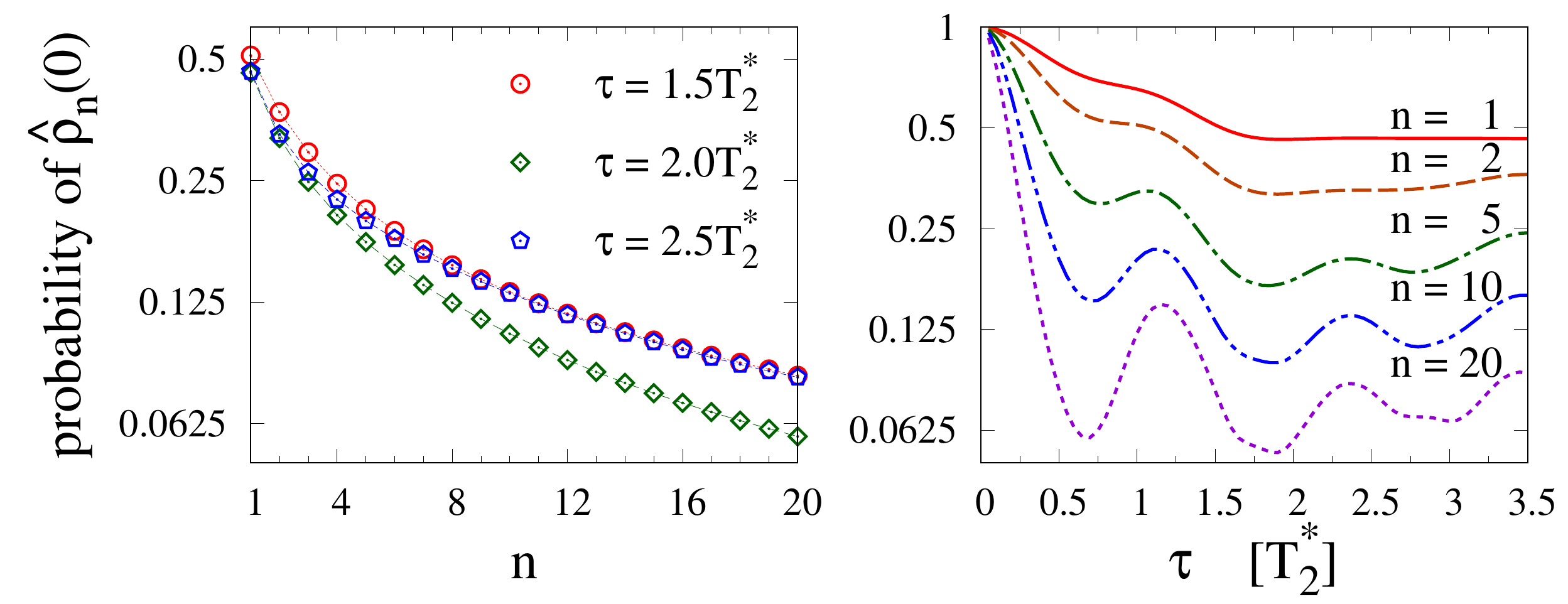}
  \caption{
            Map:
            Probability to obtain the state $\hat \rho_n(0)$ 
            as a function of number $n$ of performed projective measurements
            and time~$\tau$ between them. \\
            Graphs (cross sections of the map):
            Probability to obtain the state $\hat \rho_n(0)$ 
            as a function of number $n$ of performed projective measurements (left panel)
            and 
            as a function of time~$\tau$ between projective measurements (right panel). 
            NSEs consist of $N_1=N_2=5$ uniformly coupled spins~$\frac{1}{2}$. 
            The system is in moderate magnetic field 
            $\Omega = 5 \left[ \frac{\hbar}{T_2^*} \right]$.
          }
  \label{fig:probability_plot}
\end{figure}

\begin{figure}[t!]
  \centering
  \includegraphics[width=\linewidth, viewport=0 0 432 360, clip]{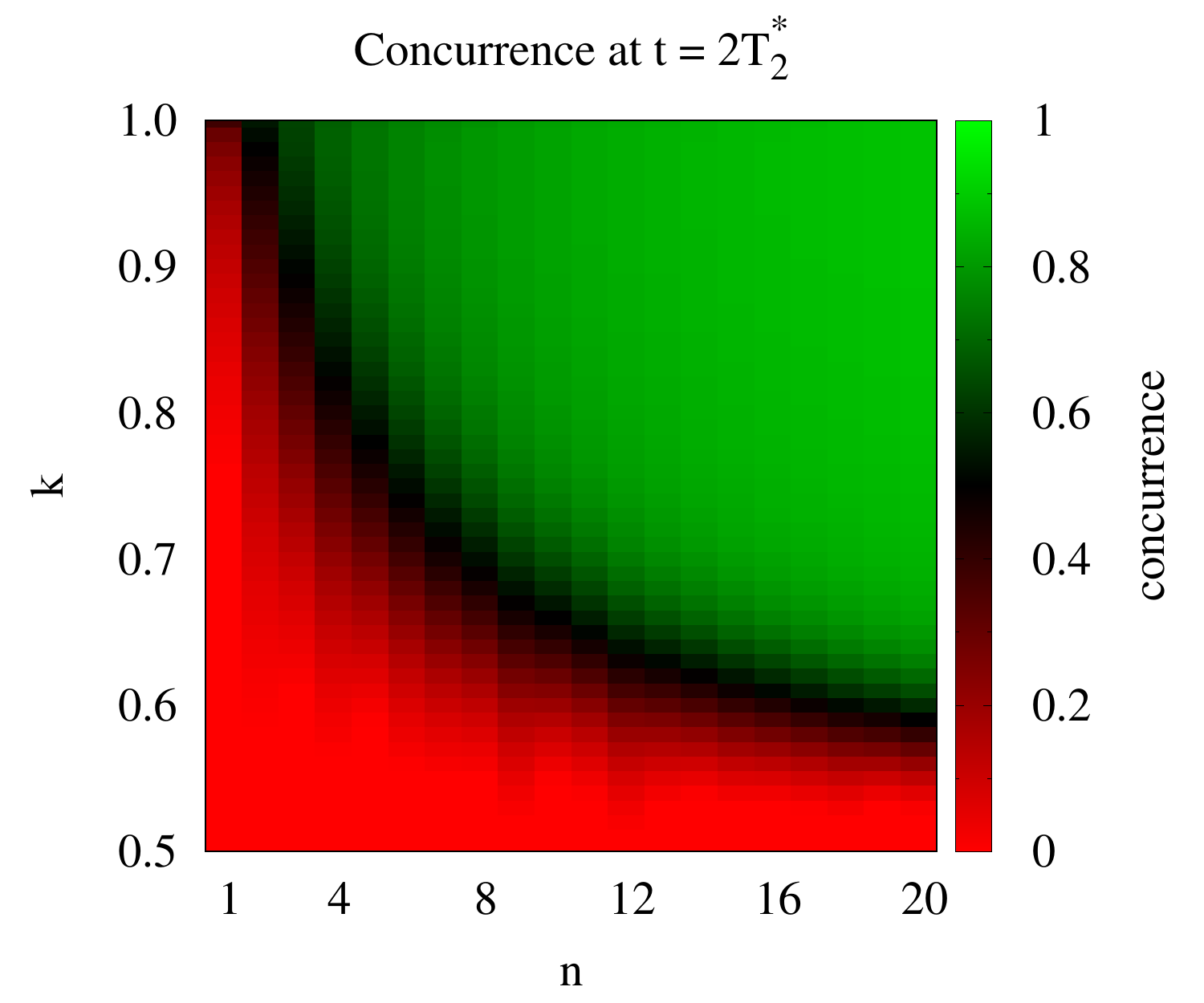}
  \includegraphics[width=\linewidth, viewport=0 0 360 252, clip]{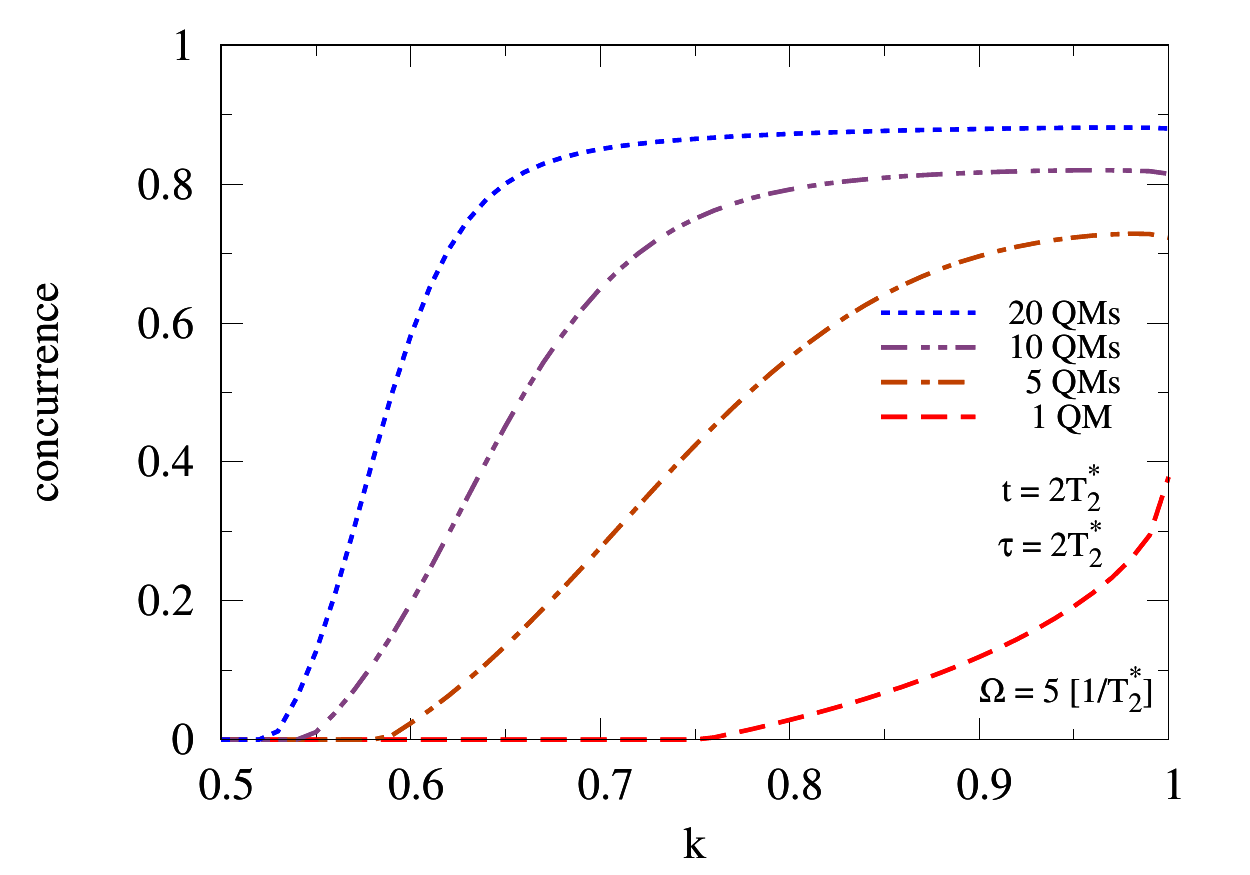}
  \caption{
            Map:
            Concurrence of two-qubit state calculated at~$t = 2T_2^*$
            as a function of strength $k$ of~quantum measurements (QM)
            performed with~period~$\tau = 2T_2^*$.\\
            Graph (cross section of the map): 
            Concurrence of two-qubit state calculated at~$t = 2T_2^*$
            as a function of strength $k$ of~quantum measurements 
            performed with~period~$\tau = 2T_2^*$.
            NSEs consist of $N_1=N_2=5$ spins~$\frac{1}{2}$. 
            The system is in moderate magnetic field 
            $\Omega = 5 \left[ \frac{\hbar}{T_2^*} \right]$.
          }
  \label{fig:concurrence_k_plot}
\end{figure}

The probability to obtain the desired state $\h \rho_n(0)$,
which is shown in left panel of Fig.~\ref{fig:probability_plot}, 
decreases monotonically 
with number $n$ of performed cycles 
due to the fact that in each cycle
the probability to obtain the postselected TESSS state
which is the same as the initial one 
is strongly less than one.
It is also worth noting 
that probability of $\h \rho_n(0)$ decreases sub-exponentially 
with increasing $n$, 
so it drops relatively slow 
and it is at the level a few percent after execution of about $n = 20$ cycles.

For $\tau > T_2^*$,
probability of $\h \rho_n(0)$,
which is shown in right panel of Fig.~\ref{fig:probability_plot},
becomes a weakly dependent function of $\tau$ 
for a fixed parameter $n$
and oscillates around the corresponding mean value.

For practical purposes,
one should find the optimal combination of procedure parameters $n$ and $\tau$
such that maximizes simultaneously 
the effect of retardation of entanglement decay (see Fig.~\ref{fig:concurrence_map})
and the probability to obtain such a state (see map in  Fig.~\ref{fig:probability_plot}).

In Fig.~\ref{fig:concurrence_k_plot} 
the dependence of intensity of the effect 
on strength $k$ of quantum measurement used in the manipulation procedure is shown.
It turns out 
that with increasing parameter $n$
concurrence of TESSS state, as a function of $k$, 
gradually develops a plateau at nearly maximal possible for a given value of $n$ level.
The plateau is situated between $k = 1$ and some lower value of $k$,
and for increasing parameter $n$, progressively reaches surprisingly low values of $k$.
Thus, a long sequences of manipulation cycles lower requirement 
for the strength $k$ of quantum measurement with practically no loss in the end effect.

The possibility to significantly retard entanglement decay 
by performing 
the manipulations with quantum measurements
originates from the
non-Markovian dynamics of the system
\cite{Coish_PhysRevB_2004,Ferraro_PhysRevB_2008}.
During free evolution, 
electron spins, initially being in an entangled state,
transfer through the hyperfine interaction
some of their quantum correlations to their nuclear spin environments.
Execution of the quantum measurement of two-electron spin subsystem
with subsequent postselection of two-electron spin state 
restores its quantum correlations. 
The state of nuclear spin subsystems, conditioned on past dynamics, in turn, 
preserves the previously obtained from electron spins quantum correlations,
and thus, the flow rate of quantum correlations 
from electron spins to nuclear spin environments
in following periods of the system evolution may be reduced,
which is manifested as the retardation of electron spin entanglement decay.

\section{Conclusions} 
\label{sec:Conclusions} 
In contrast to the fast decay of TESSS entanglement 
on a timescale of the order of  $ T^*_2$
(shown in Fig.~1 of~\cite{Mazurek_PhysRevA_2014} or~\cite{Bragar_PhysRevB_2015} 
and here in Fig.~\ref{fig:concurrence_negativity_plot}),
performing a few cycles of evolution of initially entangled two electron spin qubits
interacting with their NSEs
followed by quantum measurement performed on TESSS
gradually builds up coherences in the entire system
and the rest decay of quantum correlations of TESSS may be significantly slowed down
for specific cycle durations $\tau$
and numbers $n$ of the performed cycles.

The disadvantage of such a way of counteracting the decoherence 
is the necessity to postselect the proper two-qubit state after each quantum measurement 
and the associated with that decreasing probability of success.
On the other hand, 
the probability to obtain the desired state $\hat \rho_n(t)$ decreases sub-exponentially 
with $n$.

The strong (projective) measurements produce 
maximal effect of retardation of entanglement decay, 
but the effect can be also achieved in the case of weak measurements.
The more cycles have been performed (the larger $n$), 
the weaker quantum measurements can be used to achieve a nearly maximal effect.

Since the proposed procedure of retardation of entanglement decay 
requires only the execution of quantum measurements of two-electron subsystem,
its practical realization 
seems to be much easier 
than execution of dynamical decoupling of qubits from their environments
or preparation of a narrowed nuclear spin bath state,
but due to the indeterminicity involved in the manipulation procedure,
only a fraction of executed runs will give the desired state $\hat \rho_n(t)$,
and therefore it is not the most convenient way to counteract the decoherence.
On the other hand, 
simulations show
that when one applies the manipulation procedure with a number of cycles $n \geqslant 10$, 
the quantum measurement need not to be of projective type ($k = 1$)  anymore,
it can be of moderate strength ($k \approx 0.8$),
and the probability to obtain the desired state $\hat \rho_n(t)$, 
which will exhibit a slower decay of entanglement,
is pretty large (about 10\%). 
Thus, it may be viewed of fundamental interest to implement such a manipulation procedure 
in currently existing systems of two-electron spin QD qubits 
in order to check experimentally 
whether predicted effect of retardation of entanglement decay 
is achievable in real systems.

\bibliographystyle{unsrturl}
\bibliography{QD_qubits}

\end{document}